\documentclass{jfm}
\usepackage{amssymb,latexsym,bm,color,amsmath,natbib,epsfig}

\newcommand{\ez}{{\mathbf {\hat e}_{z}}}

\newcommand{\Cvec}{\mathbf{C}}
\newcommand{\uvec}{{\mathbf u}}
\newcommand{\xvec}{{\mathbf x}}
\newcommand{\avec}{{\mathbf a}}

\newcommand{\chivec}{{\boldsymbol \chi}}

\newcommand{\kvec}{{\mathbf k}}

\newcommand{\Uvec}{{\mathbf U}}
\newcommand{\vvec}{{\mathbf v}}
\newcommand{\Lmat}{{\mathcal L}}
\newcommand{\Nmat}{{\mathcal N}}
\newcommand{\Pmat}{{\mathcal P}}
\newcommand{\Smat}{{\mathcal S}}
\newcommand{\Ovec}{\mathbf{\Omega}}
\newcommand{\Pvec}{\mathbf{\Pi}}
\newcommand{\zerovec}{{\mathbf 0}}

\newcommand{\dotp}[2]{  {#1}\cdot {#2}  }
\newcommand{\rem}[1]{}
\newcommand{\remfig}[1]{#1}

\begin{document}
\title[Multi-frequency Craik-Criminale solutions]
   {Multi-frequency Craik-Criminale solutions of the Navier-Stokes
equations}
\author[Bruce R. Fabijonas and Darryl D. Holm]%
{B\ls R\ls U\ls C\ls E\ns R.\ns F\ls A\ls B\ls I\ls J\ls O\ls N\ls A\ls S$^1$%
 \ns \and
 \ns D\ls A\ls R\ls R\ls Y\ls L\ns D.\ns H\ls O\ls L\ls M$^{2,3}$}

\affiliation{$^1$Department of Mathematics, 
	Southern Methodist University, Dallas, TX 75275, USA\\[\affilskip]
$^2$Center for Nonlinear Studies and Theoretical Division, 
	Los Alamos National Laboratory, Los Alamos, NM  87545, USA\\[\affilskip] 
$^3$Mathematics Department,
Imperial College of Science, Technology and Medicine,
London SW7 2AZ, United Kingdom}

\pubyear{200x}
\volume{xxx}
\pagerange{xxx-xxx}
\date{\today~and in revised form ??}
\setcounter{page}{1}

\maketitle
\begin{abstract}
An exact Craik-Criminale (CC) solution to the incompressible
Navier-Stokes (NS) equations describes the instability of an elliptical
columnar flow interacting with a single Kelvin wave. These CC
solutions are extended to allow multi-harmonic Kelvin waves to interact
with any exact ``base'' solution of the NS equations. 
The interaction is evaluated along an arbitrarily chosen
flowline of the base solution, so exact nonlinear instability in this
context is locally
convective, rather than absolute.  Furthermore, an iterative
method called ``WKB-bootstrapping'' is introduced which successively adds
Kelvin waves with incommensurate phases to the extended CC solutions. This
is illustrated by constructing an extended CC solution consisting of
several Kelvin waves with incommensurate phases interacting with an
elliptical columnar flow.
\end{abstract}

\section{Introduction}
\cite{craik:crim:86} showed formally that the
sum of a base flow that is linear in the spatial coordinates and a single
traveling wave is a global solution of the incompressible Navier-Stokes (NS)
equations on an unbounded spatial domain.  Although this fact was
known in certain specific cases (see, for example, \cite[pp. 85-86]{chandra:61}),
it was Craik \& Criminale who formalized the general framework.
This discovery provided 
an exact nonlinear interpretation of elliptic instability, the
resonant mechanism
by which vortex stretching creates three-dimensional  instabilities in
swirling two-dimensional flows \cite[]{bayly:86}. This nonlinear interpretation of elliptic
instability also holds when additional physics such as buoyancy
\cite[]{miya:fuku:92}, rotation
\cite[]{craik:89,miya:93,bay:holm:lif:96} and magnetic fields
\cite[]{craik:88}
are included.  Most recently, elliptic instability has been used to
study the effects of turbulence models
\cite[]{cambon:etal:94,fab:holm:03a,fab:holm:03b}.  
Elliptic
instability and its interpretation using the Craik-Criminale (CC)
solutions were recently reviewed by  \cite{kers:02}.

A second equally ground-breaking body of work related to the CC class
of solutions is Lifschitz's WKB-like stability method for fluid mechanics
known as the Geometrical Optics Stability Method 
(GOSM) \cite[]{lif:ham:91,lif:94}.  The
method linearizes the perturbation equations about the unperturbed flow
and examines the evolution of a high-frequency infinitesimal wavepacket in
a Lagrangian reference frame which moves with the unperturbed flow.  The
power of this method is its insensitivity to the unperturbed flow's global
geometry and time dependence.  Remarkably, the  equations governing the
nonzero leading order terms of GOSM coincide with  those for CC
solutions.  This coincidence  motivates the present paper.

This paper bridges the gap between GOSM and CC solutions by extending the
CC class of exact nonlinear solutions of the NS equations in two stages.
First, we show that the CC construction method works when a Kelvin wave
together with any number of its harmonics interacts with {\it any} exact
solution of the nonlinear NS equations,
provided the interaction is evaluated along some arbitrarily
chosen flowline of the base solution.  This extension of the CC
construction method changes its interpretation from absolute instability to
locally
convective instability \cite[]{hue:mon:90}.  Second, we introduce an iterative
method called ``WKB-bootstrapping'' which constructs extended CC solutions
interacting with additional traveling waves whose phases are incommensurate.
These are multi-frequency CC solutions.  (The phrase ``incommensurate
phases'' will be explicitly defined in Eq.~\eqref{eq:IC-def}.)
The WKB-bootstrapping algorithm allows Kelvin waves with
incommensurate phases to be added to the base flow {\it one at a time}.
The power of this method is illustrated by deriving the exact
solution for nonlinear instability of an elliptical columnar flow
that is interacting with several traveling Kelvin waves whose phases are
incommensurate. Also, a steady circular columnar flow interacting with a
single Kelvin wave is found to be critically stable. That is, either
perturbing the streamline to introduce eccentricity, or adding multiple
traveling waves with incommensurate phases will yield Kelvin waves with
exponentially growing amplitudes.

\section{Craik-Criminale solutions for generalized base flows}
We consider the NS equations in a coordinate system rotating about the
$z-$axis with angular velocity $\Ovec = \Omega \ez$:
\begin{eqnarray}\label{eq:NS}
\partial_t\uvec + \dotp{\uvec}{\nabla}\uvec + 2\Ovec\times \uvec
     + \nabla p = \nu\Delta\uvec
\,.
\end{eqnarray}
Here $\dotp{\nabla}{\uvec} = 0$, where $\uvec$ is a velocity field,
$p$ is the pressure, and $\nu$ is the kinematic viscosity.  Let $\Uvec$
be any exact solution to Eq.~\eqref{eq:NS} with corresponding pressure $P$.
Consider a new exact solution of the form
$\uvec = \Uvec + \Uvec'$, $p = P + P'$.
In the analysis of nonlinear instability to follow, we shall
refer to $\Uvec$ as the `base' flow and $\Uvec'$ as the  `disturbance.'
We proceed by inserting this additive velocity decomposition into
Eq.~\eqref{eq:NS}, and using the condition that $\Uvec$ is an exact
solution, together with the vector identity
\begin{eqnarray}
(\dotp{\uvec}{\nabla})\vvec = (\nabla\times\vvec)\times\uvec
         + (\nabla\vvec)^T\cdot\uvec
\,.
\end{eqnarray}
Consequently, the nonlinear disturbance dynamics is governed by
\begin{eqnarray}\label{eq:dist}
\partial_t\Uvec' + (\dotp{\Uvec}{\nabla})\Uvec'
         + (\nabla\Uvec)^T\cdot\Uvec'
     + (\dotp{\Uvec'}{\nabla})\Uvec' + \Pvec\times\Uvec'+
     \nabla P' = \nu\Delta\Uvec',
\end{eqnarray}
and $\dotp{\nabla}{\Uvec'} = 0$.  Here,
$\Pvec = 2\Ovec + \nabla\times\Uvec$ denotes the total vorticity of
the base flow.
We choose the disturbance to
consist of a traveling wave and its harmonics of the form
\begin{eqnarray}
\Uvec' =  \sum_{m=-\infty}^\infty
\mu^{|m|}\avec_m(t)e^{im\beta\Phi} ,  \label{eq:Udist} \qquad
P' =
\sum_{m=-\infty}^\infty i\mu^{|m|}p_m(t)e^{im\beta\Phi} .
\end{eqnarray}
Here, $\Phi = \Phi(t,\xvec)$ is the fundamental phase of the disturbance.
In linear analysis, the sum $\Uvec + \Uvec'$ would be a solution to the
NS equations linearized about $\Uvec$; that is, Eq.~\eqref{eq:dist} with
the term quadratic in $\Uvec'$  removed.  
Craik \& Criminale showed that if
$\nabla\Uvec$ is a function of time only, then the sum $\Uvec +
\Uvec'$ is an {\it exact solution} to the nonlinear equations, where $\Phi$ 
is chosen to be linear in $\xvec$.
We now extend the CC solutions, by showing the sum is an exact solution to
the NS equations for any base flow $\Uvec$, when evaluated
on an arbitrarily chosen flowline of $\Uvec$.
Specifically, we consider the flowline $\xvec =
\chivec(t,\xvec_0)$ where $d\chivec/dt = \Uvec(t,\chivec)$ with
$\chivec(0,\xvec_0) = \xvec_0$, and we label this trajectory
$\mathcal{Q}_{\xvec_0}$. We emphasize that the choice of the flowline is
arbitrary. However, the subsequent equations and analysis will be
evaluated on this flowline. We seek solutions of Eq.~\eqref{eq:dist} in
the above forms whose phase $\Phi$ is frozen into the base flow.  That is,
\begin{eqnarray}\label{eq:FI}
\left ( \partial_t + \dotp{\Uvec}{\nabla} \right ) \Phi = 0,
\end{eqnarray}
with $\Phi(t,\xvec) = \kvec(t)\cdot\xvec + \delta(t)$.  Along
$\mathcal{Q}_{\xvec_0}$,
the frozen phase relation in
Eq.~\eqref{eq:FI} has the solution $\Phi(t,\chivec(t,\xvec_0)) =
\Phi(0,\xvec_0) = \Phi_0(\xvec_0)$.
The parameters
  $\mu,\beta$ are scaling factors
which allow us to choose $|\kvec(0)|=1$ and
$\max_m\{|\avec_m(0)|\} =1$.  
Evaluating the incompressibility condition $\dotp{\nabla}{\Uvec'} = 0 $
on $\mathcal{Q}_{\xvec_0}$ immediately yields
\begin{eqnarray*}
\dotp{\nabla}{\Uvec'}\Big|_{\mathcal{Q}_{\xvec_0}}
=
  \sum_{m=-\infty}^\infty
\mu^{|m|}im\beta\Big [\avec_m(t)\cdot\kvec(t)\Big ] e^{im\beta\Phi}
=0.
\end{eqnarray*}
Since the bracketed expression depends only on time along
$\mathcal{Q}_{\xvec_0}$, we conclude that
\begin{eqnarray}\label{eq:akzero}
\dotp{\avec_m}{\kvec} = 0
\end{eqnarray}
for all $m$.
Evaluating the quadratic term $\Uvec'\cdot\nabla\Uvec'$ in
Eq.~\eqref{eq:dist} along $\mathcal{Q}_{\xvec_0}$ yields
\begin{eqnarray*}
\Uvec'\cdot\nabla\Uvec'\Big|_{\mathcal{Q}_{\xvec_0}}
=
    \Bigg \{ \sum_{m=-\infty}^\infty
\mu^{|m|}\Big [\avec_m(t)\cdot\kvec(t)\Big ]e^{im\beta\Phi}\Bigg \} \Bigg
\{\sum_{n=-\infty}^\infty in\beta\mu^{|n|}\avec_n(t) e^{in\beta\Phi}\Bigg \},
\end{eqnarray*}
which also vanishes exactly. Thus, on $\mathcal{Q}_{\xvec_0}$,
the incompressibility condition $\nabla\cdot\Uvec'=0$
implies the transversality condition $\avec_m\cdot\kvec = 0$. This, in
turn, implies that the quadratic term $\Uvec'\cdot\nabla\Uvec'$ in
Eq.~\eqref{eq:dist} exactly vanishes on $\mathcal{Q}_{\xvec_0}$.
Substituting the frozen-phase solution ansatz in Eq.~\eqref{eq:FI} into
Eq.~\eqref{eq:dist} yields the Eulerian equation,
\begin{eqnarray}
\left ( \partial_t + (\nabla\Uvec)^T \right )
    \avec_m  + \Pvec\times\avec_m - m\beta p_m\nabla\Phi
    = -\nu m^2\beta^2|\nabla\Phi|^2\avec_m
\, .\label{eq:pream}
\end{eqnarray}
At this point, it is clear that we  
can take $\avec_m$ and $p_m$ to be real valued functions without loss of 
generality.  Furthermore, it follows that both the real and imaginary
parts of Eq.~\eqref{eq:Udist} individually are solutions.  
We may further decompose the base flow as $\Uvec(\xvec,t) =
\bar\Uvec(\xvec,t) + \Cvec(t)$, where $\Uvec(\xvec_0,t) = \Cvec(t)$, and set $d_t\delta + \Cvec\cdot\kvec =0$,
where $\delta(0) = \Phi_0(\xvec_0) - \kvec(0)\cdot\xvec_0$. This relation
eliminates all  contributions of $\Cvec(t)$ to the evolution of $\Phi$
and uniquely implies $\delta(t)$.
Thus, along $\mathcal{Q}_{\xvec_0}$, Eq.~\eqref{eq:dist}
reduces to a set of ODEs.  These
are obtained by taking the gradient of
Eq.~\eqref{eq:FI}, and evaluating
Eq.~\eqref{eq:pream} along $\mathcal{Q}_{\xvec_0}$:
\begin{align}
& d_t\chivec = \Uvec 
\, ,
\\
& \left (d_t + \Lmat^T \right )\kvec = \zerovec \label{eq:phase2}
\, ,
\\
&\left ( d_t + \Lmat^T \right )
    \avec_m  + \Pvec\times\avec_m - m\beta p_m\kvec
    = -\nu m^2\beta^2|\kvec|^2\avec_m\label{eq:am}
\, .
\end{align}
Here, the quantity
\begin{eqnarray}\label{eq:gradient}
\Lmat_{ij} = (\nabla\Uvec)_{ij}\Big |_{\xvec = \chivec(t,\xvec_0)}
=\partial_jU_i\Big |_{\xvec = \chivec(t,\xvec_0)}
\end{eqnarray}
is the velocity gradient tensor along $\chivec$,
and $d_t = \partial_t + \Uvec\cdot\nabla$ is the
material derivative.
Note that, $\Lmat = \nabla\Uvec$ is a function of time only, when
evaluated along $\mathcal{Q}_{\xvec_0}$.
The operator $d_t + \Lmat^T$ comprises the total time
derivative in a Lagrangian frame that moves with the base flow $\Uvec$
along $\mathcal{Q}_{\xvec_0}$. In contrast,
\cite{craik:crim:86} considered the case of {\it global} Eulerian solutions,
rather than solutions along particular Lagrangian trajectories.   
In their framework, only
base flows of the form $\Uvec = \Lmat(t)\xvec + \Cvec(t)$ are admissible.
We have
shown, however, that along a single, arbitrarily chosen
Lagrangian trajectory, the CC construction method works for {\it any}
exact solution of the NS equations. Our result reduces to Craik
\& Criminale's in the case when global Eulerian solutions are required.

We rewrite the pressure $p_m$ in Eq.~\eqref{eq:am} as follows.
Upon taking the dot product of Eq.~\eqref{eq:am} with $\kvec$ and noting
that $\dotp{\avec_m}{\kvec}$ is an integral of motion, we
solve for the pressure contribution $p_m$ of the
$m$-th harmonic in terms of the amplitude $\avec_m$
and the wavevector $\kvec$, as
\begin{eqnarray}\label{eq:pressdef}
m\beta p_m|\kvec|^2 = \dotp{(\Lmat+ \Lmat^T)\avec_m}{\kvec} +
\dotp{\Pvec\times\avec_m}{\kvec}.
\end{eqnarray}
Finally, the viscosity may be removed from the right
hand side of Eq.~\eqref{eq:am} by the change of
variables,
\begin{eqnarray}\label{eq:visc}
\avec_m = \tilde\avec_m\exp
\left ( -\int_0^t \nu m^2\beta^2|\kvec(\tau)|^2~d\tau \right )
\,.
\end{eqnarray}
One then solves for $\tilde\avec_m$ via Eq.~\eqref{eq:am} with the
right  hand side replaced by the zero vector.

We emphasize that the construction method
presented here is based on three assumptions:
(1) the disturbance amplitudes $\avec_m$ and $p_m$
are independent of the spatial variable;
(2) $\Phi$ is linear in $\xvec = \chivec(t,\xvec_0)$
along a flowline in the Lagrangian frame of the base flow $\Uvec$; and
(3) all of the analysis is carried out along a single, arbitrarily chosen
flowline specified by its initial position $\xvec_0$.
The construction method presented here fails if  $\avec_m$ is allowed
to have spatial dependence.  Such spatial dependence may be admitted by
linearizing the equations of motion about $\Uvec$ and introducing high
frequency wavepackets  rather than traditional traveling waves
\cite[]{lif:ham:91,lif:94,gjaja:holm:96}.  We note that the
superposition of traveling wave harmonics {\it does not alter the
evolution of the base flow}.  This is because the base flow $\Uvec$ is
an exact solution, by itself, whose evolution drives the disturbance in
Eq.~\eqref{eq:dist}.

\section{WKB bootstrapping}
The general framework outlined in the previous section 
allows for the construction of multi-frequency CC
solutions.  We call this method `WKB-bootstrapping' and outline an
iterative algorithm for applying it. Let us denote iterations of the
WKB-bootstrapping method with a parenthetical superscript.
Beginning with an exact base flow $\Uvec = \uvec^{(0)}$ for $n=0$, we
iteratively construct
\begin{eqnarray}\hspace{-5mm}
\uvec^{(n+1)} \!\!= \uvec^{(n)}
\!\!+\!\!
\sum_m(\mu^{(n+1)})^m\avec_m^{(n+1)}\sin(m\beta^{(n+1)}\Phi^{(n+1)}),
\end{eqnarray}
where $\Phi^{(n+1)} = \Phi^{(n+1)}(t,\xvec)$,
by following the process described above in
Section~2.  Specifically, at iteration $n+1$, we take $\Uvec = \uvec^{(n)}$ 
and $\Uvec'$ is the new traveling wave.  
The evolution of the solution at iteration $n+1$
is carried out along a single flowline
in a Lagrangian frame which moves with $\uvec^{(n)}$.
That is, the phase at iteration $n+1$ is $\Phi^{(n+1)}(t,\xvec)=
\kvec^{(n+1)}(t)\cdot\boldsymbol\xvec + \delta^{(n+1)}(t)$, where $\xvec =
\chivec^{(n+1)}(t,\xvec_0)$ with $\chivec^{(n+1)}(0,\xvec_0) = \xvec_0$
and
$d\boldsymbol\chi^{(n+1)}/dt = \uvec^{(n)}(t,\boldsymbol\chi^{(n+1)})$.
Thus, the velocity gradient tensor $\Lmat^{(n+1)}=\nabla\uvec^{(n)}$,
evaluated at $\xvec=\boldsymbol\chi^{(n+1)}(t,\xvec_0)$, is again a
function of time only, along the trajectory
$\mathcal{Q}^{(n+1)}_{\xvec_0}$.
Though at each iteration, the trajectory
$\mathcal{Q}^{(n+1)}_{\xvec_0}$ will differ from trajectories of
previous iterations, it is crucial that all iterations pass through the
same point
$\xvec_0$, that is, $\chivec^{(j)}(0,\xvec_0) = \xvec_0, j=1,2,\ldots,n+1$.
The construction method allows the phase $\Phi^{(n+1)}$ of the new Kelvin
wave to be incommensurate with the previous phases, by which we mean that 
the phases are not functionally related for all $\xvec$ and $t$.  That
is to say, 
\begin{eqnarray}\label{eq:IC-def}
\Phi^{(n+1)}(t,\xvec)\neq f(\Phi^{(j)}(t,\xvec)), \qquad
j=1,2,\ldots,n
\end{eqnarray}
for all $\xvec$, $t$ and any function $f$. 
Alternatively, the above 
definition for incommensurate phases can be rewritten as
\begin{eqnarray}\label{eq:IC-def-2}
\kvec^{(n+1)}(t)\neq f'(\Phi^{(j)})\kvec^{(j)}(t), \qquad
j=1,2,\ldots,n
\end{eqnarray}
for all time.  That is to say, none of the wave vectors are always 
parallel, although some wave vectors may be parallel at certain
instants in time.  

Because $\Phi^{(n+1)}$ and all of the $\Phi^{(j)}$ evolve along different
trajectories, the phases will tend to be incommensurate.
This iterative process may continue {\it ad infinitum}.
We emphasize that the traveling wave at iteration $n+1$
is affected by all the preceding waves, that is, those from
iterations $1,2,\ldots,n$.  However, the wave at iteration $n+1$ does
not influence the previous waves.

We emphasize that it is crucial that the waves in the WKB-bootstrapping
algorithm be added {\it one at a time}. 
For example, the solution 
$\uvec^{(3)} = \uvec^{(0)} + \mu^{(1)}\avec^{(1)}\sin(\beta^{(1)}\Phi^{(1)}) 
  + \mu^{(2)}\avec^{(2)}\sin(\beta^{(2)}\Phi^{(2)}) 
  + \mu^{(3)}\avec^{(3)}\sin(\beta^{(3)}\Phi^{(3)})$ 
is not in general an exact global solution of the NS equations.  
In particular, one cannot conclude that Eq.~\eqref{eq:akzero} holds
for the waves individually.  
The construction method outlined in Section~2
relies on the fact that we build the solution in three steps.  
First, we construct 
$\uvec^{(1)} = \uvec^{(0)}+\mu^{(1)}\avec^{(1)}\sin(\beta^{(1)}\Phi^{(1)})$ 
by setting $\Uvec = \uvec^{(0)}$ and 
$\Uvec' =  \mu^{(1)}\avec^{(1)}\sin(\beta^{(1)}\Phi^{(1)})$ 
into the theory of Section~2.  
We are now confident 
that $\uvec^{(1)}$ is 
an exact solution in a Lagrangian frame of 
an arbitrarily chosen flowline of $\uvec^{(0)}$ which passes 
through some point $\xvec_0$.  Next, we 
construct $\uvec^{(2)} = 
   \uvec^{(0)} + \mu^{(1)}\avec^{(1)}\sin(\beta^{(1)}\Phi^{(1)}) 
  + \mu^{(2)}\avec^{(2)}\sin(\beta^{(2)}\Phi^{(2)})$ by setting 
$\Uvec = \uvec^{(1)}$ 
and $\Uvec' = \mu^{(2)}\avec^{(2)}\sin(\beta^{(2)}\Phi^{(2)})$ into
the same theory.   Since $\Uvec$ is an exact solution, it is removed
from the problem for the evolution of $\avec^{(2)}$ and $\kvec^{(2)}$.
Then, the desired solution is an exact solution along the 
flowline in the Lagrangian frame 
of $\uvec^{(1)}$
which passes through
$\xvec_0$.  Finally, we construct the desired solution $\uvec^{(3)}$ 
by setting 
$\Uvec = \uvec^{(2)}$ and 
$\Uvec' = \mu^{(3)}\avec^{(3)}\sin(\beta^{(3)}\Phi^{(3)})$ into the
same theory.  This 
final solution is an exact solution in a Lagrangian frame of the
flowline of $\uvec^{(2)}$ which passes 
through $\xvec_0$.  
The individual flowlines are illustrated in Fig.~\ref{fig:lines}.
\remfig{%
\begin{figure}
\begin{center}
\epsfig{file=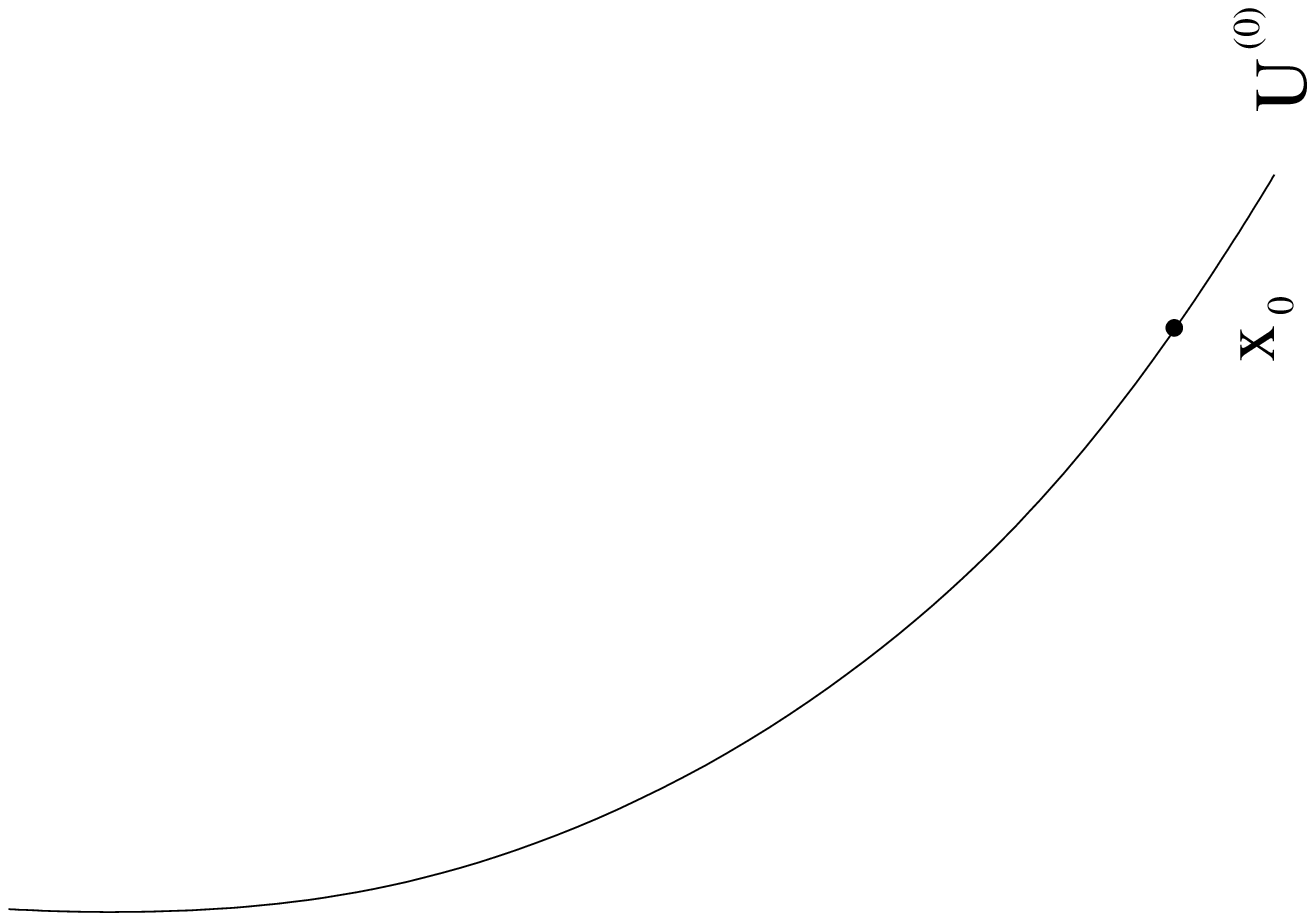,width=1.75in,angle=-90}
\epsfig{file=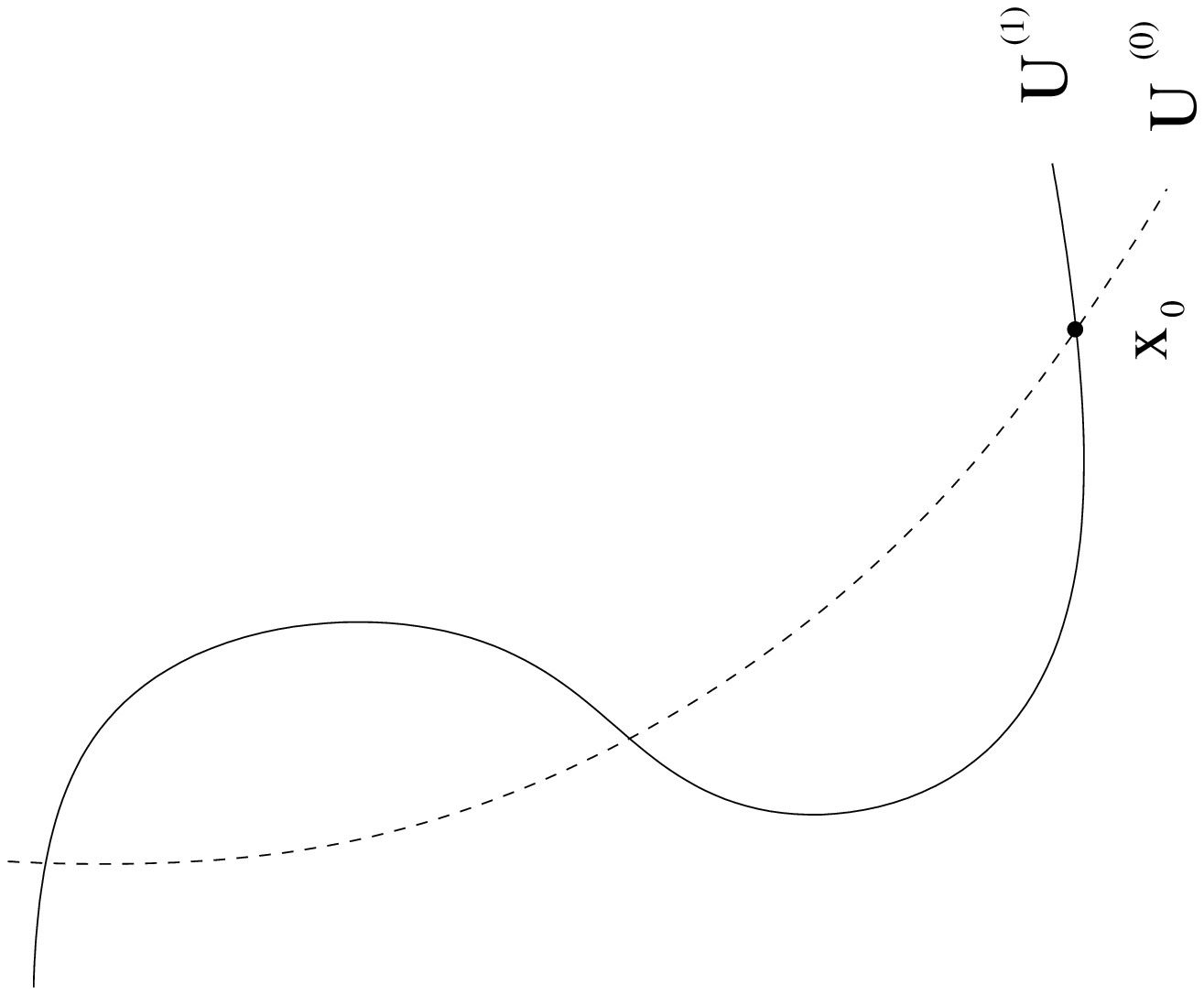,width=1.75in,angle=-90}
\epsfig{file=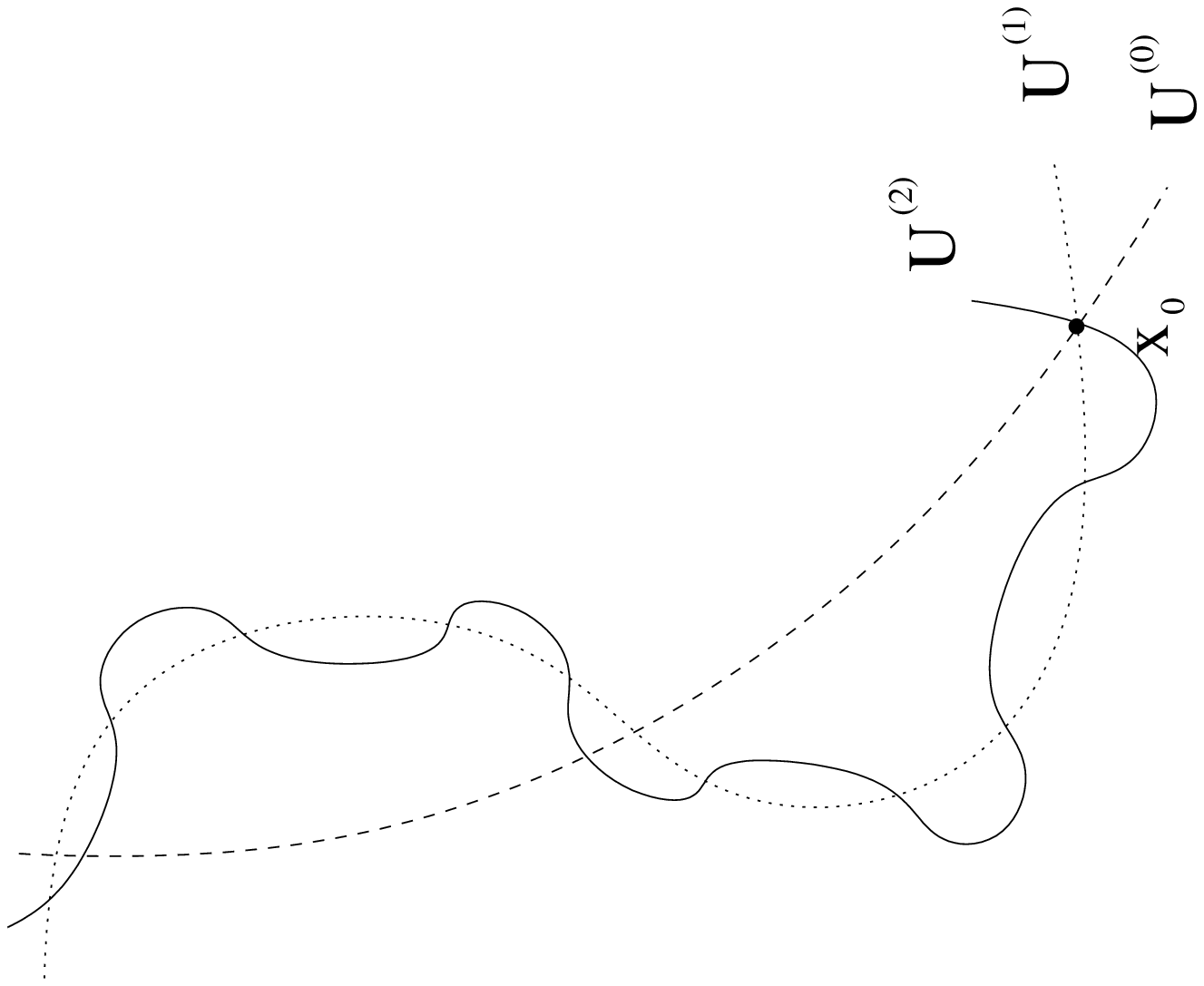,width=1.75in,angle=-90}
\caption{Representative graphic of the WKB-bootstrapping algorithm.
First, $\uvec^{(1)} = \uvec^{(0)} +
\mu^{(1)}\avec^{(1)}\sin(\beta^{(1)}\Phi^{(1)})$ 
is an exact solution in the Lagrangian frame with moves 
along an arbitrarily chosen flowline of $\uvec^{(0)}$ (left figure).
Second, $\uvec^{(2)} = \uvec^{(0)} 
  + \mu^{(1)}\avec^{(1)}\sin(\beta^{(1)}\Phi^{(1)}) 
  + \mu^{(2)}\avec^{(2)}\sin(\beta^{(2)}\Phi^{(2)})$ is an exact
solution in a Lagrangian frame which moves along a flowline of
$\uvec^{(1)}$ which passes through 
$\xvec_0$ (solid line in center figure).  Finally, 
$\uvec^{(3)} = \uvec^{(0)} 
  + \mu^{(1)}\avec^{(1)}\sin(\beta^{(1)}\Phi^{(1)}) 
  + \mu^{(2)}\avec^{(2)}\sin(\beta^{(2)}\Phi^{(2)}) 
  + \mu^{(3)}\avec^{(3)}\sin(\beta^{(3)}\Phi^{(3)})$ 
is an exact solution in a Lagrangian frame which moves along a
flowline of $\uvec^{(2)}$ which passes
through $\xvec_0$ (solid line in the right figure).
\label{fig:lines}
}
\end{center}
\end{figure}
} 
Finally, we note that by adding the waves one at a time in different
Lagrangian frames, we generate the transversality conditions
$\avec^{(1)}\cdot\kvec^{(1)} = 0$, $\avec^{(2)}\cdot\kvec^{(2)} = 0$,
and $\avec^{(3)}\cdot\kvec^{(3)} = 0$ needed in the CC construction
method.  These sequential incompressibility relations enable the
successful application of WKB-bootstrapping, by eliminating the
quadratically nonlinear term in the evolution equation for each
frequency. 

\section{Example:  stability of a circular columnar flow}  
We illustrate the WKB-bootstrapping method with an example.

\subsection{Primary instability}
Consider a base flow of
the form $\uvec^{(0)} = \Smat\xvec$ whose streamlines are ellipses with
eccentricity $\gamma$ in a non-rotating coordinate system ($\Omega = 0$):
\begin{eqnarray}
\Smat = \begin{pmatrix} 0 & -1+\gamma & 0 \\ 1+\gamma & 0 & 0
\\ 0 & 0 & 0 \end{pmatrix}.
\end{eqnarray}
We first construct a classical CC flow of the form
$\uvec^{(1)} = \Smat\xvec + \mu^{(1)} \avec^{(1)}\sin(\beta^{(1)}\Phi^{(1)})$.
This corresponds to the classic problem of elliptic instability which was
investigated by \cite{kelvin} for the circular case
($\gamma = 0$) and by \cite{bayly:86} for the elliptic case
($\gamma > 0$)
almost a century later.
The equation for the wave vector $\kvec^{(1)}$ has the analytical solution
$
\kvec^{(1)} = [ \sin\theta\cos\hat{t} ,\kappa\sin\theta\sin\hat{t},
\cos\theta]^T,
$
where $\theta$ is polar angle the wave vector makes with the axis
of rotation, $\kappa^2 = (1-\gamma)/(1+\gamma)$, and $\hat t =
t\sqrt{1-\gamma^2}$.
The equation for the amplitude
$\avec^{(1)}$ with $\Lmat^{(1)} = \Smat$ in turn
satisfies a Floquet problem \cite[]{yaku:star:76}.
That is, Eq.~\eqref{eq:am} can be written as  $d_t\avec^{(1)} =
\Nmat(t)\avec^{(1)}$ where $\Nmat(t+\tau) = \Nmat(t)$ 
and $\tau = 2\pi/\sqrt{1-\gamma^2}$.
Thus, Lyapunov-like growth rates are determined by computing the
monodromy matrix $\Pmat$, which is the fundamental solution matrix
with identity initial condition evaluated at $t=\tau$. The growth rates are
given by
$\sigma = \ln \{\max_i|\Re (\rho_i)|\}/\tau$,
where $\rho_i, i=1,2,3$ are the eigenvalues
of $\Pmat$.  \cite{bayly:86} was able to show that
for certain values of $\gamma$ and $\theta$, the amplitude could grow
exponentially in time.  See the left figure in Fig.~\ref{fig:instabdoms}.
The linearized perturbation analysis (about $\uvec^{(0)} = \Smat\xvec$)
performed by  \cite{bayly:86} concluded that the base flow
$\uvec^{(0)}$ was unstable for all $\gamma > 0$.
The work of \cite{craik:crim:86} showed that Bayly's analysis 
was an exact solution.
\remfig{%
\begin{figure}
\begin{center}
\epsfig{file=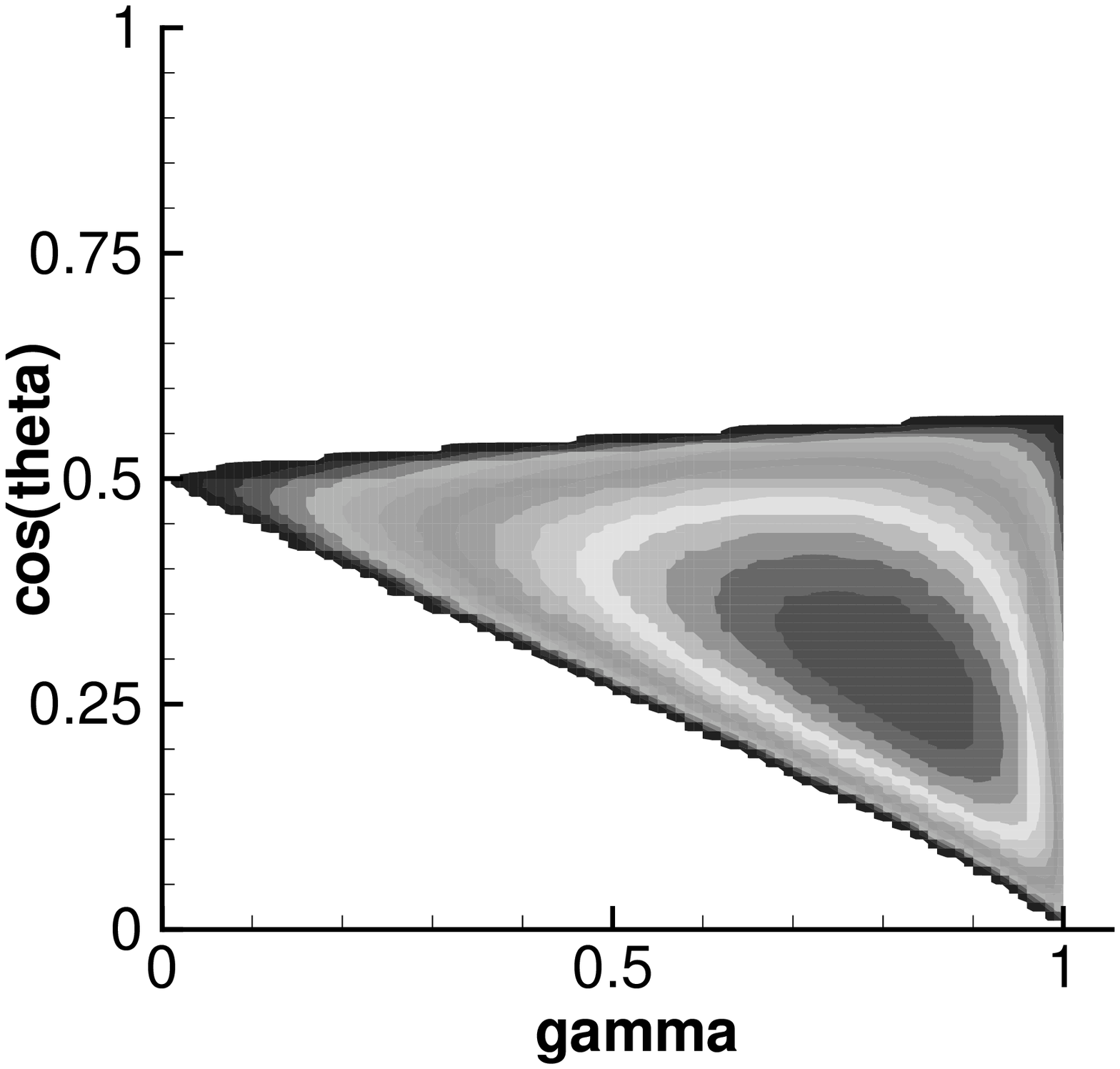,width=1.5in}
\epsfig{file=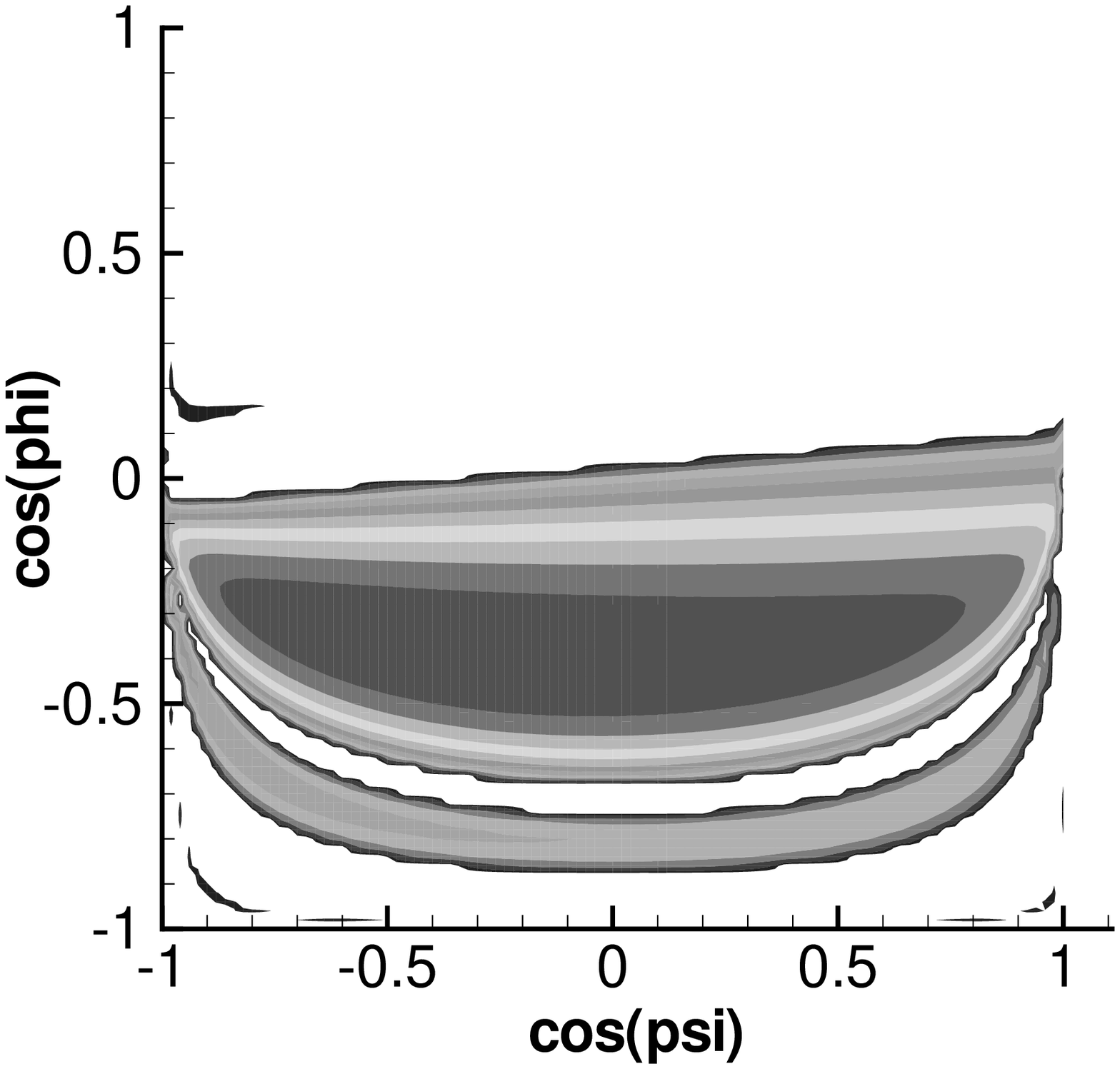,width=1.5in}
\epsfig{file=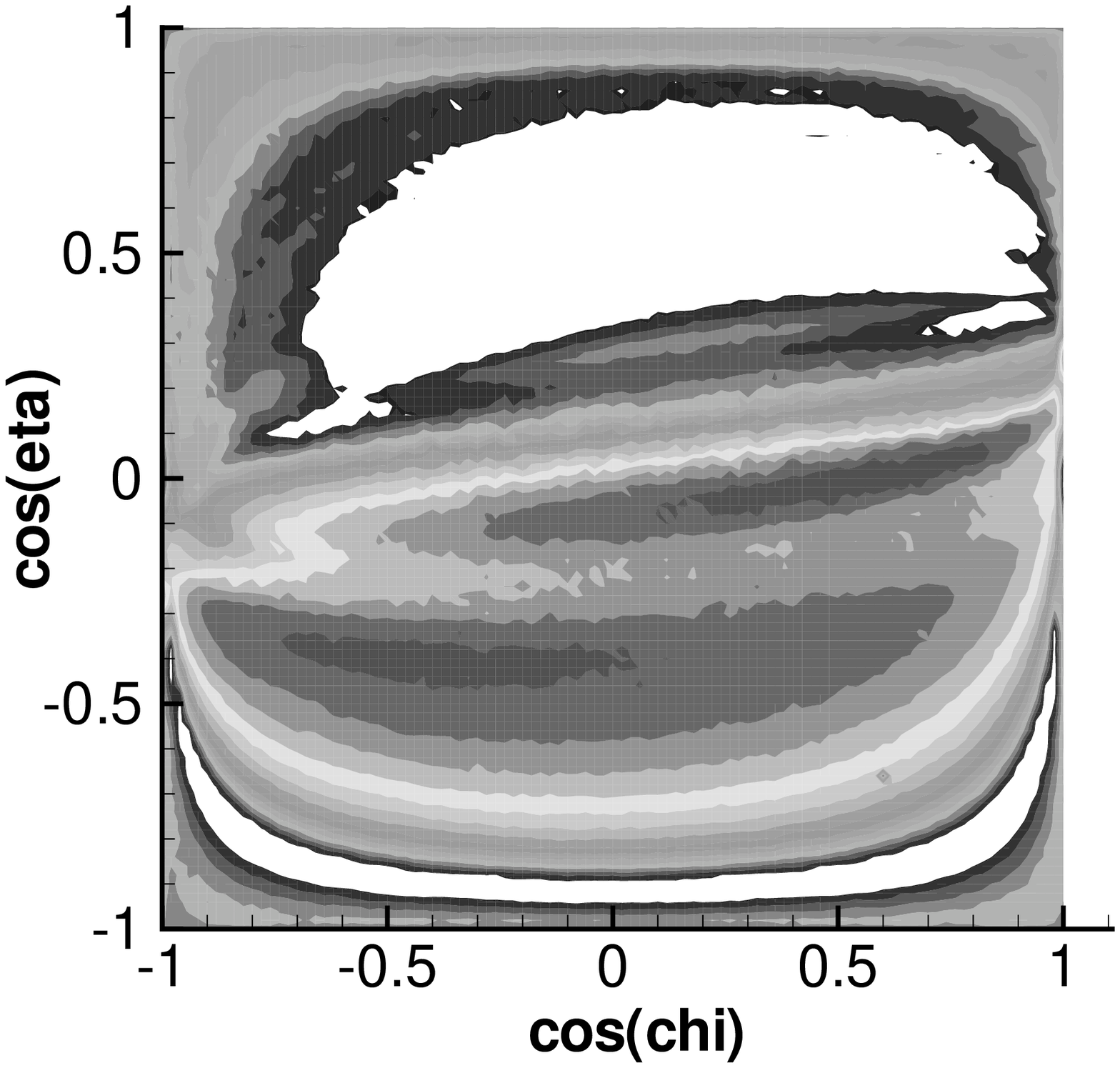,width=1.5in}
\caption{Left most figure:  
Initial condition parameter plane for $\uvec^{(1)}$.
For $\gamma = 0$, the Kelvin wave is periodic in time; for values of
$\gamma,\cos\theta$ which fall into the
white region for $\gamma > 0$, the Kelvin wave is quasiperiodic, that
is, the periods of $\avec^{(1)}$ and $\kvec^{(1)}$ are
rationally related; for the remaining region, the Kelvin wave has an
exponentially growing amplitude.
Center figure:  
Initial orientation parameter plane for $\uvec^{(2)}$ at $\xvec =
\zerovec$ with parameter values $\cos\theta = 0.2$, $\gamma = 0$
and  $\mu^{(1)}\beta^{(1)} = 1$.
For parameter values which fall into the white regions, the amplitude
of $\avec^{(2)}$ is bounded; for all others, the amplitude of $\avec^{(2)}$
grows exponentially in time.
Right figure:
Initial orientation parameter plane for $\uvec^{(3)}$ at $\xvec =
\zerovec$ with parameter values $\cos\theta = 0.2$, $\gamma = 0$,
$\cos\phi = 0.5$, $\cos\psi = 0$.
For parameter values which fall into the white regions, the amplitude
of $\avec^{(3)}$ is bounded; for all others, the amplitude of $\avec^{(3)}$
grows exponentially in time.
\label{fig:instabdoms}
}
\end{center}
\end{figure}
} 

\subsection{Secondary instability}
We now construct the next WKB-bootstrap iteration as $\uvec^{(2)} =
\Smat\xvec + \mu^{(1)}\avec^{(1)}\sin(\beta^{(1)}\Phi^{(1)}) +
\mu^{(2)}\avec^{(2)}\sin(\beta^{(2)}\Phi^{(2)})$.
The equations for  the phase
$\Phi^{(2)}$ and the amplitude $\avec^{(2)}$ are given
in Eqs.~\eqref{eq:phase2}-\eqref{eq:pressdef} with $\Uvec =
\uvec^{(1)}$.  In the interest of simplicity, we consider the case
$\cos\theta = 0.2, \gamma = 0$.   Then $\uvec^{(1)}$ is periodic with
period $2\pi$,  and the equation for $\avec^{(2)}$ again
satisfies a Floquet problem.  We
further simplify our analysis by examining the flow at the origin,
which is a fixed point of the system.
Then, $\Phi^{(1)} = 0$ and
$\Lmat^{(2)} = \Smat +
\mu^{(1)}\beta^{(1)}\avec^{(1)}(\kvec^{(1)})^T$.
Note that although $\Phi^{(1)} = 0$ and $\Phi^{(2)} = 0$ {\it at this fixed
point}, this does not imply that the phases are commensurate.  Indeed, 
we choose the initial condition for the wave vector $\kvec^{(2)}$ to
be arbitrary in three-dimensions: 
$\kvec^{(2)}(0) = [ \cos\psi\sin\phi, \sin\psi\sin\phi,
\cos\phi ]^T$.  The center figure in Fig.~\ref{fig:instabdoms}
shows the domain in the $(\cos\phi,\cos\psi)$
plane for which the amplitude of $\avec^{(2)}$ grows exponentially.
This figure first appeared in \cite{lif:fab:96} in which the
authors examined the stability of Bayly's solution for $\gamma = 0$
using GOSM.
We conclude, then that the analysis of \cite{lif:fab:96} is also
{\it an exact solution of the full nonlinear Euler equations}.  This result
applies to the subsequent articles by the same authors and their
collaborators \cite[]{fab:holm:lif:97,fab:lif:98}.
The authors viewed the perturbation analysis of $\uvec^{(1)}$ as a
secondary perturbation analysis of $\uvec^{(0)}$.

\subsection{Tertiary instability}
We construct a {\it third iteration} of this method as $\uvec^{(3)}
= \Smat\xvec + \mu^{(1)}\avec^{(1)}\sin(\beta^{(1)}\Phi^{(1)}) +
\mu^{(2)}\avec^{(2)}\sin(\beta^{(2)}\Phi^{(2)}) +
\mu^{(3)}\avec^{(3)}\sin(\beta^{(3)}\Phi^{(3)})$,
where the phases are incommensurate.
Again, we examine the case $\cos\theta = 0.2$, $\gamma = 0$ at
the fixed point $\xvec
= \zerovec$, and additionally we choose $\cos\phi = 0.5$ and $\cos\psi =
0$.  Thus, the flow $\uvec^{(2)}$ is bounded in time, and
$\Lmat^{(3)} = \Smat + \mu^{(1)}\beta^{(1)}\avec^{(1)}(\kvec^{(1)})^T +
\mu^{(2)}\beta^{(2)}\avec^{(2)}(\kvec^{(2)})^T$.  The individual
components of $\uvec^{(2)}$ are periodic, but there is no apparent
common period.  To analyze the evolution of
$\avec^{(3)}$, we use a quasi-periodic analogue of Floquet theory
as discussed in detail in \cite{bay:holm:lif:96}.
We use the incompressibility condition $\dotp{\avec^{(3)}}{\kvec^{(3)}} =
0$ to reduce Eq.~\eqref{eq:am} for $\avec^{(3)}$ to a two-dimensional
system. We then use the Pr\"ufer transformation
\begin{eqnarray}
a_1^{(3)}(t) = e^{d(t)}\cos[c(t)], \quad a_2^{(3)}(t) = e^{d(t)}\sin[d(t)]
\,,
\end{eqnarray}
where $a_1^{(3)}$ and $a_2^{(3)}$ are the first two
components of $\avec^{(3)}$.
This produces a coupled pair of ordinary differential equations for
$c(t)$ and $d(t)$.  Note that linear growth of $d(t)$ implies
exponential growth for $\avec^{(3)}$.  One can define two quantities
called the growth rate $I$ and the winding number $W$ as
\begin{eqnarray}
I = \lim_{t\to\infty} [d(t)/t], \quad
W = \lim_{t\to\infty} [c(t)/t],
\end{eqnarray}
respectively.  It was first shown conclusively in
\cite{bay:holm:lif:96} that these quantities my be approximated
by long time simulations independently of the values
of $c(0)$ and $d(0)$.  We choose the initial condition for
$\kvec^{(3)}$ in polar coordinates as before:
$\kvec^{(3)}(0) = [\cos\chi\sin\eta,\sin\chi\sin\eta,\cos\eta]^T$.
The right figure in Fig.~\ref{fig:instabdoms} 
shows the parameter plane $(\cos\eta,\cos\chi)$
for which the amplitude $\avec^{(3)}$ grows exponentially in time for
a specific orientation of $\avec^{(2)}$.
In the spirit of
\cite{lif:fab:96}, we may view this as a {\it tertiary}
perturbation of the rigidly rotating flow which again
provides an exact nonlinear NS solution.

\section{Discussion}
The present analysis extends the previous results for
stability of a rigidly rotating column of fluid with circular
streamlines.  \cite{bayly:86} showed that though this case was
stable in the presence of a traveling wave, slight
perturbations in the streamline eccentricity
yield a traveling wave whose amplitude grows exponentially in time
(left figure in Fig.~\ref{fig:instabdoms}).
Furthermore, \cite{lif:fab:96} showed that while the
primary traveling wave was always stable, the amplitude of a secondary wave
with an incommensurate phase grows exponentially (center figure in 
Fig.~\ref{fig:instabdoms}).  In the present paper, we show that even
when the secondary wave is stable (that is, then the amplitude is 
bounded for all time), the tertiary wave is unstable for a 
larger portion of its parameter space.  Thus, we conclude that 
a circular columnar flow is critically stable in the sense that
either perturbations in the eccentricity or the addition of
incommensurate phases will result in waves whose amplitudes grow
exponentially in time.  

This paper has accomplished three main objectives.  First,
we extended the pioneering work of \cite{craik:crim:86} and
\cite{bayly:86} 
to allow nonlinear instability analysis of any exact
solution of the NS equations due to multi-harmonic Kelvin waves,
evaluated along a flowline in the Lagrangian frame of the base flow.
Second, we showed how to construct a sequence of exact
solutions to the NS equations that involves the multi-frequency
superposition of Kelvin traveling waves by adding the waves one at a
time.  This iterative method of
constructing exact NS solutions was called `WKB-bootstrapping.'
Third, we used the WKB-bootstrapping method to unite and extend various
classical linearized instability analyses of a circular columnar flow.

\acknowledgements
The authors are indebted to A. Lifschitz-Lipton whose work has
guided us throughout this investigation. We are also grateful
to anonymous referees whose comments significantly improved the
presentation of the material.
BF thanks the Theoretical Division at the Los Alamos National
Laboratory for their hospitality and acknowledges valuable exchanges with
members of the Laboratory's Turbulence Working Group.


\end{document}